\documentclass[12pt]{article}
\pdfoutput=1
\usepackage{graphicx}
\usepackage{amssymb}
\usepackage{amsmath}
\usepackage{natbib}
\newcommand{\db}{{\bf x}}

\newcommand{\dbobs}{\db_o}

\newcommand{\ssb}{\textrm{\boldmath$s$}}
\newcommand{\ssbobs}{\ssb_o}

\newcommand{\phibn}{\textrm{\boldmath$\phi$}}
\newcommand{\Pb}{P_{\theta}}
\newcommand{\Pbsub}{P}
\newcommand{\Qb}{Q_{\theta}}
\newcommand{\Qbsub}{Q}
\newcommand{\Zb}{\mathcal{Z}_{n,m}}

\begin{document}

\title{AABC: approximate approximate Bayesian computation when simulating a large number of data sets is computationally infeasible}
     % Approximate Bayesian inference using small number of simulated data sets under the model
\author{Erkan O. Buzbas  \\
Department of Biology \\ Stanford University, Stanford, CA 94305-5020 USA  \\ 
and\\
Department of Statistical Science \\ University of Idaho, Moscow, ID 84844-1104 USA\\
email: \texttt{erkanb@uidaho.edu}\\
and\\
Noah A. Rosenberg  \\
Department of Biology \\ Stanford University, Stanford, CA 94305-5020 USA  \\ 
email: \texttt{noahr@stanford.edu}
}
\maketitle

\newpage
\begin{center}
\textbf{Abstract}
\end{center}
Approximate Bayesian computation (ABC) methods perform inference on model-specific parameters of mechanistically motivated parametric statistical models
when evaluating likelihoods is difficult.
Central to the success of ABC methods is computationally inexpensive simulation of data sets from the parametric model of interest.
However, when simulating data sets from a model is so computationally expensive that the posterior distribution of parameters cannot be adequately sampled by ABC, inference is not straightforward.
We present ``approximate approximate Bayesian computation'' (AABC), a class of methods that extends simulation-based inference by ABC to models in which simulating data is expensive.
In AABC, we first simulate a \emph{limited number} of data sets that is computationally feasible to simulate from the parametric model.
We use these data sets as fixed background information to inform a non-mechanistic statistical model that approximates the correct parametric model and enables efficient simulation of a large number of data sets by Bayesian resampling methods. 
We show that under mild assumptions, the posterior distribution obtained by AABC converges to the posterior distribution obtained by ABC, as the number of data sets simulated from the parametric model and the sample size of the observed data set increase simultaneously. 
We illustrate the performance of AABC on a population-genetic model of natural selection, as well as on a model of the admixture history of hybrid populations.

\vspace*{.3in}


\newpage

\section{Introduction}\label{Intro}
%----------------------------------------------------------------
Stochastic processes motivated by mechanistic considerations enable investigators to capture salient phenomena in modeling natural systems.
Statistical models resulting from these stochastic processes are often parametric, and estimating model-specific parameters---which often have a natural interpretation---is a major aim of data analysis.
Contemporary mechanistic models tend to involve complex stochastic processes, however, and parametric statistical models resulting from these processes lead to computationally intractable likelihood functions.
When likelihood functions are computationally intractable, likelihood-based inference is a challenging problem that has received considerable attention 
in the literature \citep{RobertCasella2004, Liu2008}.
\par
%------------------------------------------------------------------------
When statistical models are known only at the level of the stochastic mechanism generating the data---such as in implicit statistical models \citep{DiggleGratton1984}---explicit evaluation of likelihoods might be impossible.
In these models, standard computational methods that require evaluation of likelihoods up to a proportionality constant (e.g., rejection methods) cannot be used to sample distributions of interest.
However, data sets simulated from the model under a range of parameter values
can be used to assess parameter likelihoods without explicit evaluation \citep{Rubin1984}.
Approximate Bayesian computation (ABC) methods \citep{Tavareetal1997,Beaumontetal2002, Marjorametal2003} 
implement this idea in a Bayesian context to sample an {\em approximate} posterior distribution of the parameters. 
Intuitively, parameter values producing simulated data sets similar to the observed data set
arise in approximate proportion to their likelihood, and hence, when weighted by prior probabilities, to their posterior probabilities.
\par
%----------------------------------------------------------------------------
\subsection{The ABC literature}
ABC methods have been based on rejection algorithms
\citep{Tavareetal1997, Beaumontetal2002, BlumFrancois2010},
Markov chain Monte Carlo \citep{Beaumont2003, Marjorametal2003, Bortotetal2007, Wegmannetal2009}, and
sequential Monte Carlo \citep{Sissonetal2007, Sissonetal2009, Beaumontetal2009, Tonietal2009}.
Model selection using ABC \citep{Pritchardetal1999, Fagundesetal2007, Grelaudetal2009,
BlumJakobsson2010, Robertetal2011}, the choice of summary statistics when the likelihood is based on 
summary statistics instead of the full data \citep{JoyceMarjoram2008, Wegmannetal2009, NunesBalding2010, FearnheadPrangle2012},
and the equivalence of posterior distributions targeted in different ABC methods \citep{Wilkinson2008, Sissonetal2010} have also 
been investigated.
\par
%--------------------------------------------------------------
ABC methods have had a considerable effect on model-based inference in disciplines that rely on genetic data, particularly data 
shaped by diverse evolutionary, demographic, and environmental forces.
Example applications have included problems in the demographic history of populations \citep{Pritchardetal1999, Francoisetal2008, Verduetal2009, BlumJakobsson2010} and species
\citep{Estoupetal2004, PlagnolTavare2004, BecquetPrzeworski2007, Fagundesetal2007, Wilkinsonetal2010},
as well as problems in the evolution of cancer cell lineages \citep{Tavare2005, Siegmundetal2008} and the evolution of protein networks \citep{Ratmannetal2009}.
Other applications outside of genetics have included inference on the physics of stereological extremes \citep{Bortotetal2007},
the ecology of tropical forests \citep{JabotChave2009},  dynamical systems in biology
\citep{Tonietal2009}, and small-world network disease models \citep{Walkeretal2010}.
ABC methods have been reviewed by \citet{MarjoramTavare2006}, \citet{Cornuetetal2008}, \citet{Beaumontetal2009}, \citet{Beaumont2010},
\citet{Csilleryetal2010}, and \citet{Marinetal2011}.
\par
%------------------------------------------------------------------
\subsection{A limitation of ABC methods}
An informal categorization of the information available about the likelihood function is helpful
to illustrate the class of models in which ABC methods are most useful.
First, exact inference on the posterior distribution of the parameters is possible only if the likelihood function is analytically available. 
Second, if the likelihood function is not analytically available but can be evaluated up to a constant given a parameter value, then standard computational methods such as rejection algorithms can sample the posterior distribution.
In this case, inference is exact up to a Monte Carlo error due to sampling from the posterior.
Third, if the likelihood function cannot be evaluated, but data sets can feasibly be simulated from the model, then ABC methods sample the posterior distribution using approximations on the {\em data space} in addition to a Monte Carlo error due to sampling.
\par
%--------------------------------------------------------------------
Although ABC methods sample the posterior distribution of parameters without evaluating the likelihood function, they are computationally intensive.
Adequately sampling a posterior distribution of a parameter by ABC requires many random realizations from the prior distribution
of the parameter and the sampling distribution of the data.
Simulating from the prior is straightforward, but the computational cost of simulating a data set from the mechanistic model increases quickly with the complexity and number of stochastic processes involved.
Henceforth, we refer to statistical models in which not only evaluating the likelihoods is difficult but also simulating a large number of data sets is computationally infeasible as {\em limited-generative} models. 
When a model is limited-generative and only a small number of data sets can be simulated from the model, likelihoods cannot be assessed using ABC and hence, the posterior distribution of parameters cannot be adequately sampled. 
\par
%--------------------------------------------------------
\subsection{Our contribution}
\begin{figure}
\begin{center}
\includegraphics[height=6.7cm]{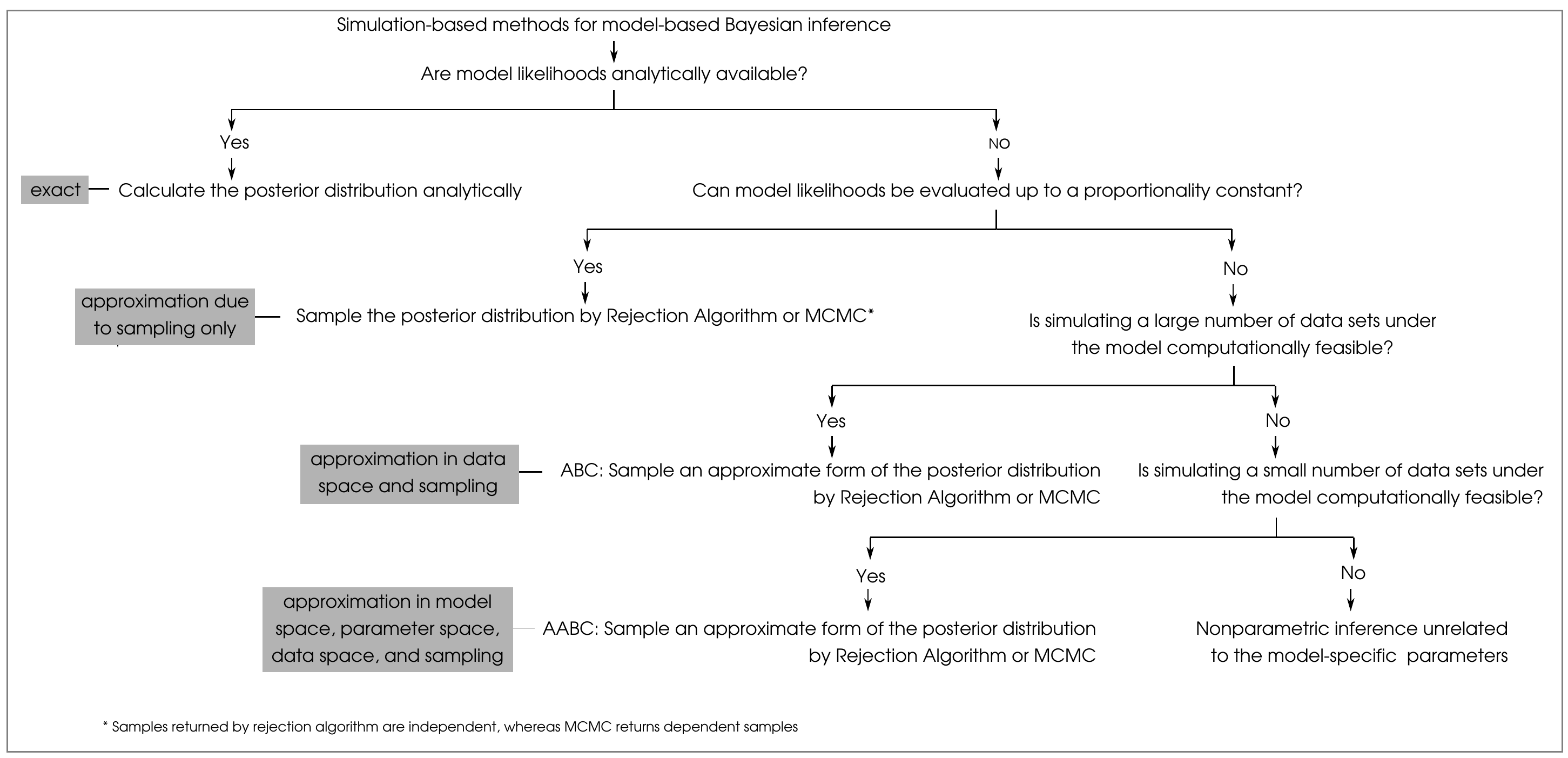}
\caption[]{Applicability of simulation-based inference methods in relation to the information available about the likelihood function.}
\label{fig:1}
\end{center}
\end{figure}

In this article, we introduce {\em approximate} approximate Bayesian computation (AABC), a class of methods that perform inference on model-specific parameters of limited-generative models when standard ABC methods are computationally infeasible to apply.
In AABC, the idea of assessing the likelihoods approximately using simulated data sets is taken one step further than in ABC.
AABC methods make approximations on the {\em parameter space} and the {\em model space} in addition to standard ABC approximations on the data space.
In conjunction with Bayesian resampling methods, these approximations help us overcome the computational intractability associated with simulating data from a limited-generative model (Figure \ref{fig:1}).
\par
%-----------------------------------------------------------------------------------------
Our key innovation is to condition on a limited number of data sets that can be feasibly simulated from the limited-generative model and to employ a non-mechanistic statistical model to simulate a large number of data sets.
We set up the non-mechanistic model based on empirical distributions of the limited number of data sets simulated from the mechanistic model.
Since the data values from the limited number of simulated data sets are used to construct new random data sets by resampling methods, it is computationally inexpensive to simulate a large number of data sets in AABC. 
The AABC approach allows a researcher to allocate a fixed computer time to simulating a limited number of data sets from the limited-generative model,
thus making otherwise challenging likelihood-based inference attainable.
\par
%-----------------------------------------------------
Intuitively, the information conditioned upon by the non-mechanistic model increases with the number of data sets simulated from the mechanistic model, and the expected accuracy of inference obtained by AABC methods increases. 
We formalize this intuition by showing that the posterior distribution of parameters obtained by AABC converges to the corresponding posterior distribution obtained by standard ABC, as the sample size of the observed data set and the number of data sets simulated from the limited-generative model increase simultaneously.
\par
%-----------------------------------------------------------------
%--------------------------------------------------------
\par
%--------------------------------------------------------
AABC methods utilize the established machinery of ABC methods in sampling the posterior distribution of the parameters.
Therefore, standard approximations on the data space involved in an ABC method---which facilitate the sampling of the posterior distribution---apply to AABC methods as well. 
We now briefly review these approximations in the context of ABC by rejection algorithms.
%
%---------------------------------------------------------------------------
\section{Review of ABC by rejection algorithms}\label{sec:ABCreview}
%-----------------------------------------------------------------
To more formally set up the class of problems in which ABC methods are useful, we assume that a parametric model generates observations conditional on parameter $\theta \in \Theta \equiv \mathbf{R}^p,\;p\geq1.$
We let $\Pb$ be the sampling distribution of a data set of $n$ observations independent and identically distributed (IID) from this model.
We denote a random data set by $\db=(x_1,x_2,...,x_n) \in \mathcal{X},$ where $\mathcal{X}$ is the space in which the data set sits, and the observed data set by $\dbobs.$
In the genetics context, a data point $x_i$
might be a vector denoting the allelic types of a genetic locus at genomic position $i$ in a group of individuals; 
the data matrix $\db$ might then contain genotypes from these individuals in a sample of $n$ independent genetic loci. 
\par
%--------------------------------------------------------------
Suppose that $\Pb$ is available to the extent that the likelihood function $p(\dbobs|\theta)$ can be evaluated up to a constant whose value does not depend on the parameters.
Given a prior distribution $\pi(\theta)$ on parameter $\theta,$ the posterior
distribution of $\theta$ given the observed data $\dbobs$ under the model $\Pb$ is 
$\pi(\theta|\dbobs, \Pb).$ 
Then $\pi(\theta|\dbobs, \Pb)$ can be sampled by standard rejection sampling from $p(\dbobs|\theta)\pi(\theta),$ a quantity that is proportional to 
$\pi(\theta|\dbobs, \Pb)$ by Bayes' Theorem.
In principle, sampling $\pi(\theta|\dbobs, \Pb)$ without evaluating the
likelihood function $p(\dbobs|\theta)$ is possible, if simulating the data from the model $\Pb$ is feasible.
An early example due to Tavar\'e {\em et al.} (1997) samples $\pi(\theta|\dbobs,\Pb)$
by accepting a value $\theta_i$ simulated from the prior $\pi(\theta)$
only if the data set $\db_i$ simulated from $\Pbsub_{\theta_i}$ satisfies $\db_i=\dbobs.$   
By standard rejection algorithm arguments, the $\theta_i$ sampled in this fashion are from the correct posterior distribution.
However, the acceptance condition $\db_i=\dbobs$ is rarely satisfied with high-dimensional data. 
A first approximation in ABC methods is dimension reduction by substituting the data set $\db$ with a low-dimensional set of summary statistics $\ssb.$
The observed data $\dbobs$ and the simulated data $\db_i$ are substituted by
$\ssbobs$ and $\ssb_i,$ calculated from their respective data sets.
This is equivalent to substituting the likelihood function of the data $p(\db|\theta)$ with the likelihood function of the summary statistics $p(\ssb|\theta).$
Since ABC is most useful in statistical models that do not admit sufficient statistics, dimension 
reduction to summary statistics often entails information loss about the parameters. 
The choice of summary statistics minimizing this information loss
is an active research area \citep{JoyceMarjoram2008, Wegmannetal2009, Robertetal2011, Aeschbacheretal2012, FearnheadPrangle2012}.
\par
%---------------------------------------------------------------------------
When the data are substituted with summary statistics, the acceptance condition $\db_i=\dbobs$ is substituted by $\ssb_i=\ssbobs,$ but exact equality may still be too stringent a condition to be satisfied with simulated data.
A second approximation in ABC is to relax the exact acceptance condition with a tolerance acceptance condition.
For example, \citet{Pritchardetal1999} used the Euclidean distance $||\cdot||$ and a small tuning parameter $\epsilon$
to accept a value $\theta_i$ from an approximate posterior distribution if the data set $\db_i$ simulated from $\Pbsub_{\theta_i}$ produced $\ssb_i$ satisfying 
\begin{equation}\label{eq:euclideandistance}
||\ssb_i-\ssbobs||=\left[\sum_{j=1}^{k}(s_{ij}-s_{oj})^2\right]^{1/2}\leq\epsilon,
\end{equation}
where $\ssb$ is a $k$-dimensional statistic, and $s_{ij}$ and $s_{oj}$ are the $j$th components of $\ssb_i$ and $\ssbobs,$ respectively 
(see also \citet{WeissVonHaeseler1998} for an application in a pure likelihood inference context).
Distance metrics other than the Euclidean distance, such as the total variation distance \citep{Tavareetal2002}, have also been used.
\par
%-------------------------------------------------------------
Substituting the binary accept/reject step in the rejection sampling by weighting $\ssb_i$ smoothly according to its distance from $\ssbobs$ using a kernel density ${\rm K}_{\epsilon}(\ssb_i,\ssbobs)$ with bandwidth $\epsilon$ leads to importance sampling \citep{Wilkinson2008}.
The tolerance condition $||\ssb_i-\ssbobs||\leq\epsilon$ in the rejection algorithm of \citet{Pritchardetal1999} then corresponds to using a uniform kernel on an $\epsilon$-ball around $\ssbobs.$  
Other approaches to kernel choice include Epanechnikov \citep{Beaumontetal2002} and Gaussian \citep{LeuenbergerWegmann2010}
kernels.
\par
\begin{figure}[h]
\begin{center}
\includegraphics[height=4.8cm]{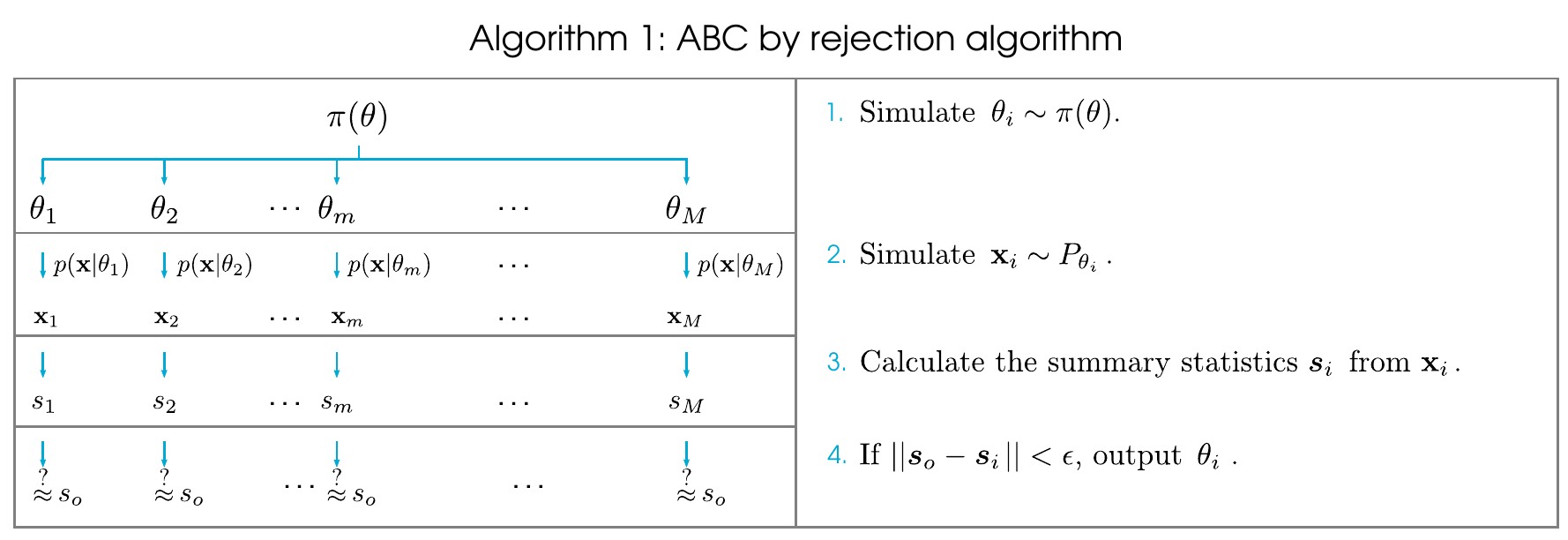}
\caption[]{The ABC algorithm by rejection sampling. One iteration of the algorithm is shown on the right along with a schematic illustration of sampling from the posterior distribution of $\theta$ based on $M$ proposed parameter values (left).}
\label{fig:2}
\end{center}
\end{figure}

%------------------------------------------------------------------
When the data likelihood is substituted by the likelihood based on the summary statistics and a
tolerance condition with a uniform kernel and the Euclidean distance is used, the posterior distribution
sampled with ABC by rejection is

\begin{equation}\label{eq:2}
\pi_{\epsilon}(\theta|\dbobs, \Pb) =\frac{1}{C_{\Pb}}\int_{\mathcal{X}} \mathbf{I}_{\{||\ssb-\ssbobs||<\epsilon\}} p(\db|\theta)\pi(\theta) \;d\db,
\end{equation}
where $\mathbf{I}_{A}$ is an indicator function that takes a value of 1 on set $A$ and is zero otherwise, 
and $C_{\Pb}=\int_{\Theta}\int_{\mathcal{X}} \mathbf{I}_{\{||\ssb-\ssbobs||<\epsilon\}} p(\db|\theta)\pi(\theta) \;d\db \;d\theta$ is the normalizing constant.
A standard ABC algorithm that samples $\pi_{\epsilon}(\theta|\dbobs, \Pb)$ appears in Figure \ref{fig:2}.
\par
%----------------------------------------------------------
%\begin{enumerate}
%\item[] \textbf{Algorithm 1 (ABC by rejection):}
%\item[] Input: $\pi(\theta),p(\db|\theta),\epsilon,||\cdot||, \dbobs.$
%\item[] Output: A draw from the posterior distribution $\pi_{\epsilon}(\theta|\dbobs,\Pb).$
%\item[1.] Simulate $\theta^* \sim \pi(\theta).$
%\item[2.] Simulate $\db^* \sim p(\db|\theta^*).$ 
%\item[3.] Calculate the summary statistics $\ssb^*$ from $\db^*.$
%\item[4.]  If $||\ssbobs-\ssb^*||<\epsilon,$ output $\theta^*$ otherwise, go to step 1. 
%\end{enumerate}
% 
%----------------------------------------------------------------------------------
%
\par
\begin{figure}
\begin{center}
\includegraphics[height=4.5cm]{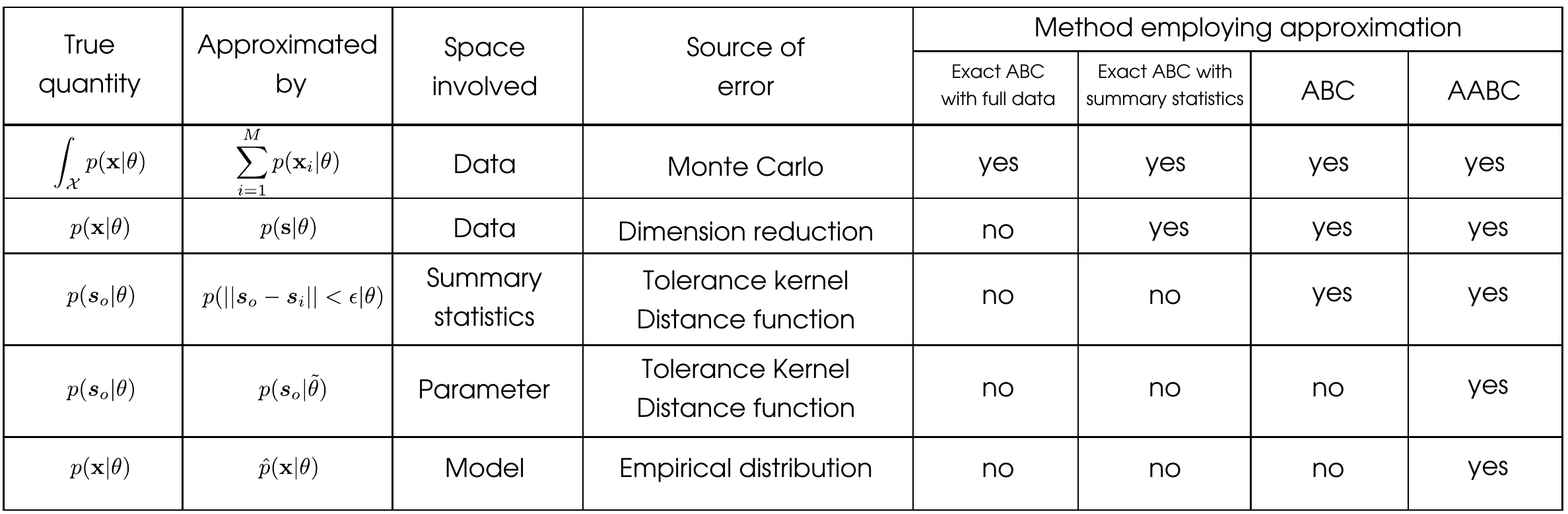}
\caption[]{Approximations and errors involved in simulation-based ABC inference methods. Likelihood functions of the full data and the summary statistics are denoted respectively by \(p(\db|\theta)\) and \(p(\ssb|\theta).\) Exact ABC with full data involves only the Monte Carlo approximation due to sampling and thus is equivalent to a standard rejection algorithm. Summary statistics \(\ssb\) are assumed not to be sufficient so that dimension reduction from \(\dbobs\) to \(\ssbobs\) results in an approximation.}
\label{tablo:1}
\end{center}
\end{figure}

%---------------------------------------------------------------------
The choice of summary statistics, tolerance parameter $\epsilon,$ distance function, and kernel constitute approximations on the data space in ABC methods.
We assume that these standard ABC approximations work reasonably well, and we focus on new modeling approximations on the parameter and model spaces introduced by AABC (Figure \ref{tablo:1}).
%---------------------------------------------------------
\par
%-------------------------------------------------------------------------------
\section{Approximate approximate Bayesian computation (AABC)}\label{sec:theory}
%------------------------------------------------------------
Algorithm 1 returns an adequate sample size from the posterior distribution of a parameter if it is iterated a large number of times, $M$.
The set of realizations simulated from the joint distribution of the parameter and the data by steps 1 and 2 of Algorithm 1 is then $\{(\db_1, \theta_1),(\db_2, \theta_2),...,(\db_M, \theta_M)\}.$
AABC methods seek inference on parameter $\theta$ when the model $\Pb$ is limited-generative, and simulating $M$ data sets under $\Pb$ is therefore computationally infeasible.
We thus assume that only a limited number $m$ of data sets $\db_1,\db_2,...,\db_m$ can be obtained by step 2 of Algorithm 1 $(m \ll M)$.
We denote the set of realizations simulated from the joint distribution of the parameter and the data by $\Zb=\{(\db_1, \theta_1),(\db_2, \theta_2),...,(\db_m, \theta_m)\},$
where each data set $\db_i$ of $n$ IID observations is simulated from the model $\Pbsub_{\theta_i}.$ 
\par
%------------------------------------------------
In AABC, we substitute the joint sampling distribution $\Pb$ of a data set of size $n$ with the joint sampling distribution $\Qb,$ from which simulating
data sets is computationally inexpensive.
In replacing $\Pb$ with $\Qb,$ we require that the posterior distribution $\pi(\theta|\dbobs, \Qb)$ based on the likelihood implied by model $\Qb$ approximates the posterior distribution $\pi(\theta|\dbobs, \Pb)$ based on the likelihood implied by model $\Pb.$
Further, we require that $\Qb$ can be used with a wide range of $\Pb,$ in the sense that $\Qb$ is constructed without using the details of model $\Pb.$ 
\par
%--------------------------------------------------------------------------
\subsection{Approximations on the parameter and model spaces due to replacing $\Pb$ with $\Qb$}\label{subsec:nonparametric}
%---------------------------------------------------------------------
Two approximations are involved in substituting $\Pb$ with $\Qb.$ 
First, $\Zb$ includes only $m$ parameter values $\theta_1,\theta_2,...,\theta_m$ under which data sets are simulated from $\Pb$. 
After obtaining $\Zb,$ for any new parameter value $\theta$ from the prior distribution under which we want to simulate a new data set, we substitute $\theta$ with $\tilde{\theta}$ such that
$(\tilde{\db},\tilde{\theta})\in\Zb.$
The value $\tilde{\theta}$ has the minimum Euclidean distance to the value $\theta$ among all parameter values in $\Zb.$
More precisely, $\tilde{\theta}=\displaystyle{\mathop{\mbox{arg\;min}}_{\theta_j \in \Zb}}||\theta_j-\theta||.$
In essence, this approximation is equivalent to replacing the sampling distribution of the data set $\Pb$ with the sampling distribution $\Pbsub_{\tilde{\theta}}$; we call this an approximation on the parameter space.
However, this parameter space approximation is not sufficient to simulate data sets efficiently, since the model $\Pbsub_{\tilde{\theta}}$ is still limited-generative after this substitution. 
\par
%--------------------------------------------------------
As a second approximation, we substitute the model $\Pbsub_{\tilde{\theta}}$ with the empirical distribution of the data set $\tilde{\db}$ that has already been simulated from 
$\Pbsub_{\tilde{\theta}}$ as $(\tilde{\db},\tilde{\theta})\in\Zb.$ 
Here, we assume a positive probability mass only on the data values observed in the set $\tilde{\db}.$
We call this an approximation on the model space because the model $\Pbsub_{\tilde{\theta}}$ is substituted with the empirical distribution of a data set simulated from 
$\Pbsub_{\tilde{\theta}}.$
\par
%---------------------------------------------------------
To simulate a new data set $\db$ in AABC, we utilize a vector of positive auxiliary parameters $\phibn=(\phi_1,\phi_2,...,\phi_{n}),$ that satisfy $\sum_{i=1}^{n}\phi_i=1.$
We let $\phi_i$ be the probability that a random data value $x_j\in\db$ is equal to a given value $\tilde{x}_i$ found in the data set $\tilde{\db}=(\tilde{x}_1,\tilde{x}_2,...,\tilde{x}_n).$ 
%
%The data set $\tilde{\db}$ is of the simulated pair $(\tilde{\db},\tilde{\theta})\in \Zb.$ 
%
The premise is that the sample $\tilde{\db}$ simulated under $\tilde{\theta}$ provides information about the model $\Pbsub_{\tilde{\theta}},$ and by an approximation of $\theta$ to $\tilde{\theta}$ on the parameter space, about $\Pb$. 
\par
%-------------------------------------------------------------
If we denote the approximate sampling distribution of a data set $\db=(x_1,x_2,...,x_n)$ by $\Qb,$ its joint probability mass function is 
\begin{equation}\label{eq:Q}
\int_{{\Phi}}q(\db|\phibn,\tilde{\db})\pi(\phibn) \;d\phibn\; \mathbf{I}_{\{\theta,\tilde{\theta}\}},
\end{equation}
where $q(\db|\phibn,\tilde{\db})={n \choose n_1 \;n_2\; \cdots\; n_k}\prod_{j=1}^{n}\prod_{i=1}^{n}\phi_i^{\mathbf{I}_{\{x_j=\tilde{x}_i\}}},$ and $\mathbf{I}_{\{\theta,\tilde{\theta}\}}$ is 1 if $\tilde{\theta}\in\Zb$ is the closest value to $\theta$ in the Euclidean sense and is 0 otherwise.  
Here, $n_i$ is the number of times $\tilde{x}_i$ observed in the new sample $\db,$ $k$ is the number of distinct data values observed in the data set $\db,$ and $\mathbf{I}_{\{x_j=\tilde{x}_i\}}$ is 1 if $x_j=\tilde{x}_i$ and is 0 otherwise.
The distribution $q(\db|\phibn,\tilde{\db})$ is that of an IID sample $\db=(x_1,x_2,...,x_n),$ where $x_j$ is drawn from the values $(\tilde{x}_1,\tilde{x}_2,...,\tilde{x}_n)$ with probabilities $(\phi_1,\phi_2,...,\phi_n).$ 
\par
%-----------------------------------------------------------------------
The probability vector $\phibn$ is a parameter of the model conditional on $\tilde{\db},$ and thus, we need to posit a prior distribution on $\phibn.$
As a natural prior on probabilities, we let the prior distribution $\pi(\phibn)$ on $\phibn$ be the symmetric Dirichlet distribution on the $(n-1)${\em-}dimensional simplex $\Phi,$ with hyperparameters (1,1,...,1) and a uniform probability density function proportional to $1.$ 
This choice assigns equal weight to all distributions placing positive probability mass on the data points $\tilde{x}_i\in\tilde{\db}.$
Further, it assigns zero posterior probability to data values unobserved in the sample $\tilde{\db},$ thereby avoiding difficulties created by such values in the likelihood \citep{Rubin1981, Owen1990}.
\par
\begin{figure}
\begin{center}
\includegraphics[height=7cm]{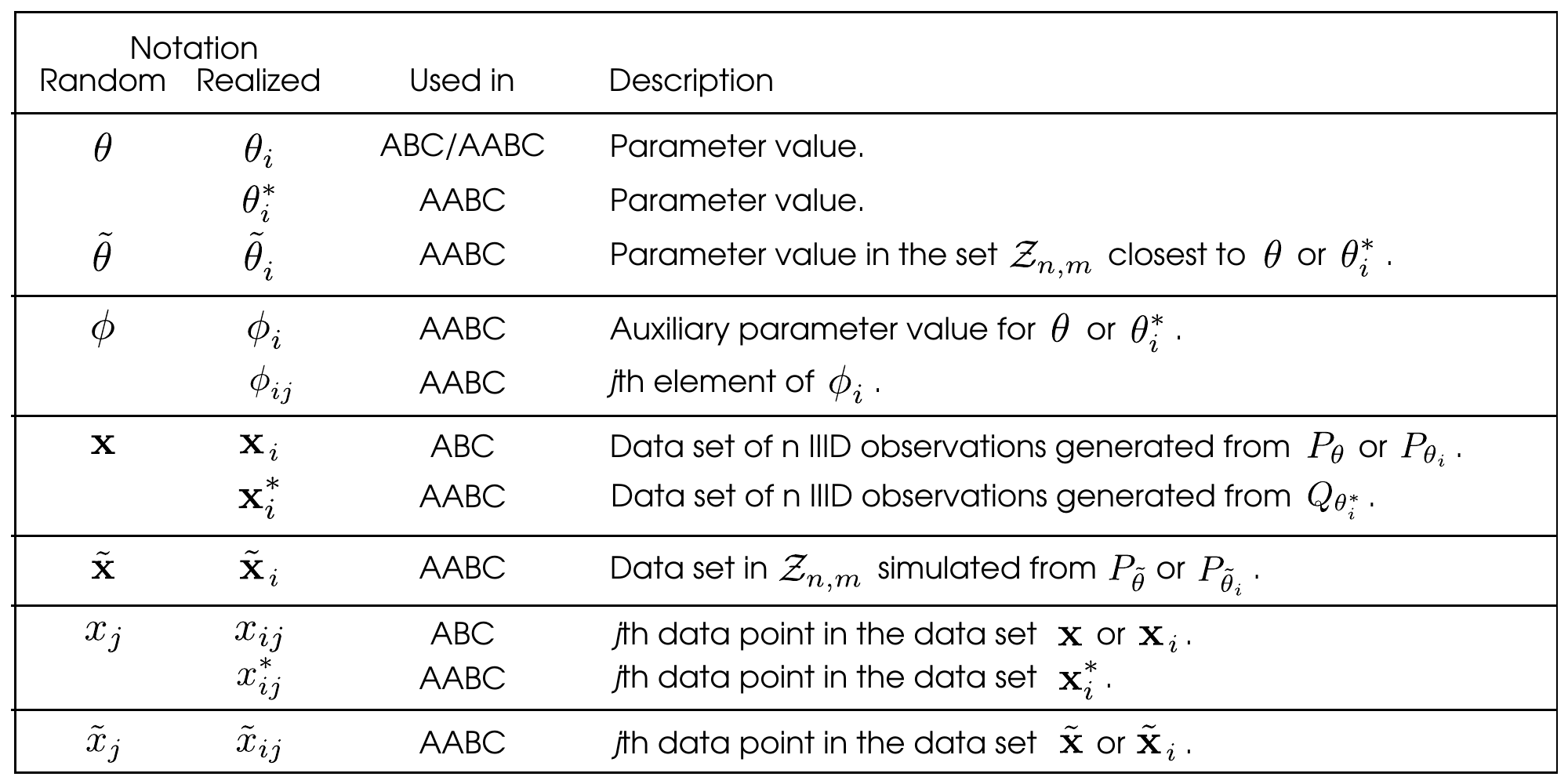}
\caption[]{Notation used in the text and algorithms.}
\label{fig:11}
\end{center}
\end{figure}

%--------------------------------------------------------------
To distinguish the parameter and data set realizations in $\Zb=\{(\db_i,\theta_i)\}_{i=1}^{m}$ from the parameter and data sets simulated using AABC, we use starred versions of each quantity to denote specific values simulated in AABC. 
For example, as the sampling distribution $\Pbsub_{\theta_i}$ delivers a data set $\db_i$ under a given parameter value $\theta_i$ in the ABC procedure of Algorithm 2, the sampling distribution 
$\Qbsub_{\theta^*_i}$ delivers a data set $\db^*_i$ under a given parameter value $\theta^*_i$ simulated from its prior distribution (see Figure \ref{fig:11} for notation). 
%-----------------------------------------------------------------
%--------------------------------------------------------------------------------
\par
%------------------------------------------------------------------------
The sampling distribution $\Qb$ utilizes the information available in the set of realizations $\Zb$ through the parameter $\phibn,$ since the prior distribution of $\phibn$ conditions on $(\tilde{\db},\tilde{\theta})\in \Zb$ and thus on the set $\Zb.$
In this sense, the available realizations $\Zb$ are used as fixed background information about $\Pb,$ and inferences using the substitute model $\Qb$ are conditional on the simulated sets $\Zb.$
\par
%------------------------------------------------------------
%------------------------------------------------------------
\subsection{The posterior distribution of $\theta$ sampled by AABC}\label{sec:posterior}
In sampling the approximate posterior distribution of $\theta$ by AABC methods, we use the two ABC approximations described in Section \ref{sec:ABCreview}.
First, we substitute each data instance $\db$ with summary statistics $\ssb.$ 
Second, we use an acceptance condition with tolerance $\epsilon,$ employing the Euclidean distance to measure the proximity of the summary statistics calculated from the observed and simulated data, as in equation \ref{eq:euclideandistance}.
If we let $\theta^*_j$ be a new parameter value simulated from its prior distribution after obtaining the set $\Zb,$ in AABC we accept
the parameter values $\theta^*_j$ producing summary statistics $\ssb^*_j$ that satisfy the condition $||\ssb^*_j-\ssbobs||<\epsilon$ as being draws from the posterior distribution.  
This acceptance condition corresponds to a uniform kernel, which we use throughout this article, although like ABC, AABC can employ other kernels
to obtain smooth weighting of $\ssb^*_j$ values by their distance from $\ssbobs.$
Substituting $\Pb$ with $\Qb$ involves replacing $p(\db|\theta)$ in expression \ref{eq:2} with expression \ref{eq:Q} and adjusting the normalizing constant accordingly.
The approximate posterior distribution sampled by an AABC method is
\begin{equation}\label{eq:4}
\pi_{\epsilon}(\theta|\dbobs,\Qb) =
\frac{1}{C_{\Qb}}\int_{\mathcal{X}}\mathbf{I}_{\{||\ssb-\ssbobs||<\epsilon\}}\left[\int_{\Phi}q(\db|\phibn,\tilde{\db})\pi(\phibn)\;d\phibn\;\mathbf{I}_{\{\theta,\tilde{\theta}\}}\right]\pi(\theta) \;d\db,
\end{equation}
where $C_{\Qb}=\int_{\Theta}\int_{\mathcal{X}}\mathbf{I}_{\{||\ssb-\ssbobs||<\epsilon\}}\left[\int_{\Phi}q(\db|\phibn,\tilde{\db})\pi(\phibn)\;d\phibn\;\mathbf{I}_{\{\theta,\tilde{\theta}\}}\right]\pi(\theta) \;d\db \;d\theta$ is the normalizing constant.
\par
%------------------------------------------------------------------------
The AABC approach is sensible in that as the limited generative model increasingly permits a larger number of simulated data sets, for large sample sizes the posterior distribution obtained by an AABC method approaches the same distribution as the posterior distribution obtained by an ABC method.
We codify this claim with a theorem. 
\par
\vskip 0.5cm
\noindent
%--------------------------------------
{\em Theorem.} 
Let $\pi(\theta)$ be a bounded prior on $\theta.$ Let $\pi_{\epsilon}(\theta|\dbobs,\Pb)$ and $\pi_{\epsilon}(\theta|\dbobs,\Qb)$ be the posterior distributions sampled by a standard ABC method and an AABC method, respectively. Then
\begin{equation}\label{eq:theorem}
\lim_{m \rightarrow \infty}\lim_{n\rightarrow \infty}\pi_{\epsilon}(\theta|\dbobs,\Qb)=\lim_{n \rightarrow \infty}\pi_{\epsilon}(\theta|\dbobs,\Pb).
\end{equation}
%-------------------------------------
A proof of the theorem is given in Appendix 1.
The convergence of the posterior distribution sampled by AABC is a consequence of the fact that, for each given value of $\theta,$ the sampling distribution $\int_{\Phi}q(\db|\phibn,\tilde{\db})\pi(\phibn)\;d\phibn\;\mathbf{I}_{\{\theta,\tilde{\theta}\}}$ converges to the true sampling distribution $p(\db|\theta)$ as the sample size $n$ and the number of simulated samples $m$ from $\Pb$ increase.
The intuition for the double limit in equation \ref{eq:theorem} is as follows.
The standard notion of a distibution converging to a point in the parameter space as the sample size $n$ increases does not directly apply to the posterior distribution $\pi_{\epsilon}(\theta|\dbobs,\Qb),$ since this posterior depends not only on the sample size $n,$ but also on the number $m$ of simulated data sets from $\Pb.$
Hence, for convergence of the posterior distribution based on the likelihood of $\Qb,$ the requirement is that both $n\rightarrow \infty$ and $m\rightarrow \infty.$
As $n\rightarrow\infty,$ the empirical distribution converges to $\Pbsub_{\tilde{\theta}},$ the correct sampling distribution with the incorrect parameter value $\tilde{\theta}.$
As $m\rightarrow \infty,$ the distance between the parameter value $\theta$ under which we want to simulate a new data set and the parameter value $\tilde{\theta}\in\Zb$ closest to $\theta$ approaches zero.
Therefore, taking both limits simultaneously results in convergence to the correct sampling distribution $\Pb.$
\par
%-------------------------------------------------------------------
\subsection{AABC algorithms}\label{subsec:ABCapproximations}
\begin{figure}[h]
\begin{center}
\includegraphics[height=8.1cm]{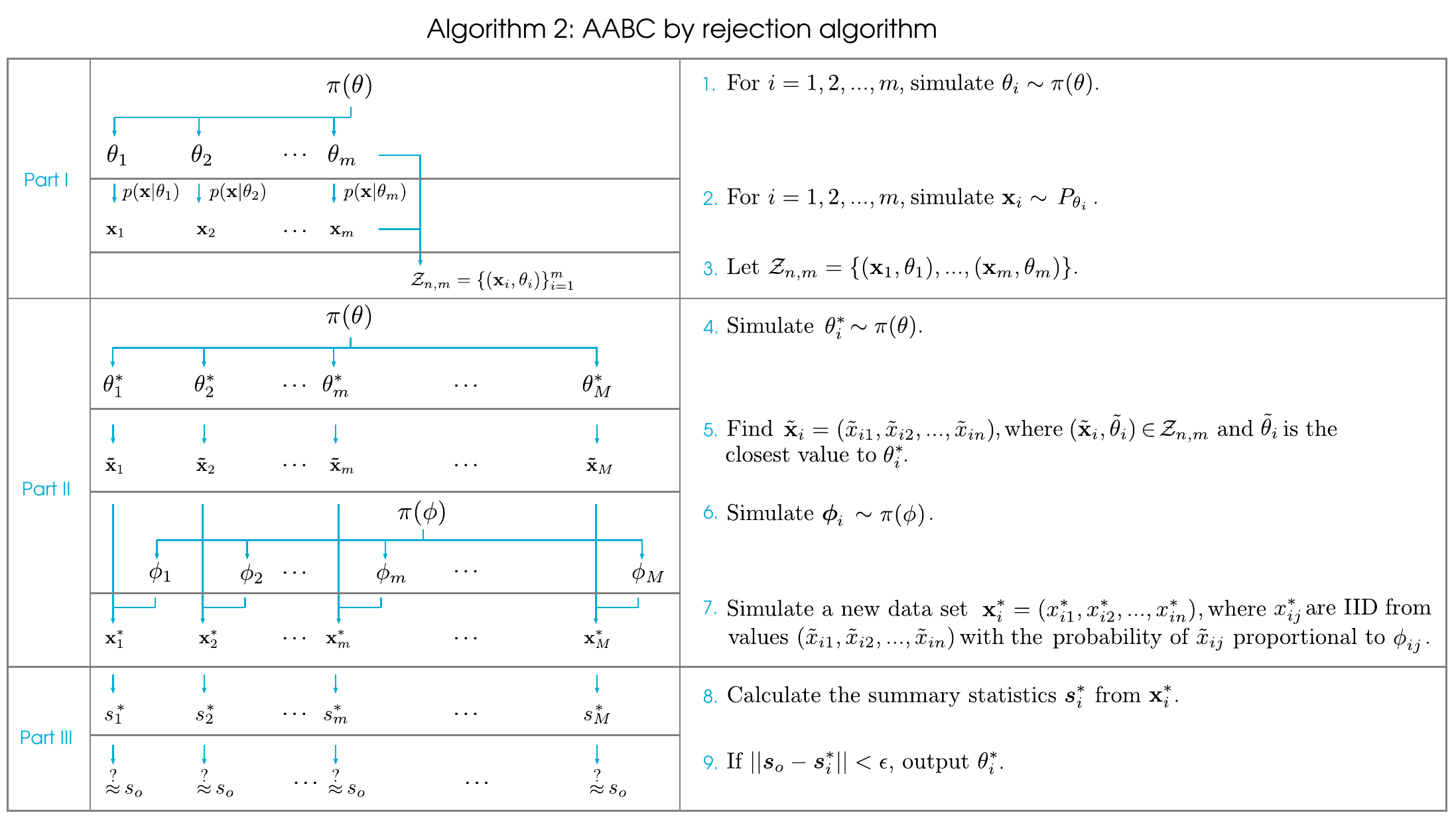}
\caption[]{The AABC algorithm by rejection sampling. One iteration of the algorithm is shown on the right, along with a schematic illustration of sampling from the posterior distribution of $\theta$ based on $M$ proposed parameter values in the rejection algorithm (left).}
\label{fig:3}
\end{center}
\end{figure}

%-------------------------------------------------------------------
The structure of AABC algorithms sampling the posterior distribution in expression \ref{eq:4} can be conveniently summarized in three parts, as shown in AABC by a rejection algorithm (Figure \ref{fig:3}).
In Algorithm 2, Part I involves obtaining a limited number of realizations 
from the joint distribution of the parameter and the data from the limited-generative model $\Pb.$ 
Part I simply involves the application of steps 1 and 2 from Algorithm 1, but only for $m$ iterations.
Part II involves simulating a new parameter value $\theta^*_i$ from its prior distribution (step 4) and then simulating a data set $\db^*_i$
from the model $\Qbsub_{\theta^*_i}$ (steps 5, 6, 7), conditional on $\Zb$ obtained in Part I.
Part III involves comparing the summary statistics $\ssb^*_i$ calculated from the simulated data set $\db^*_i$ with the summary statistics $\ssbobs$ calculated from the observed data set $\dbobs,$ to accept or reject the parameter value $\theta^*_i.$
The calculation and comparison of summary statistics follows the same procedure as in steps 3 and 4 of Algorithm 1. 
Hence, Part II of AABC by rejection has the novel steps 5, 6, and 7, whereas Parts I and III use the machinery of ABC by rejection from Algorithm 1.
\par
%------------------------------------------------------
We can show that Algorithm 2 samples the correct posterior distribution $\pi_\epsilon(\theta|\dbobs,\Qb).$
The probability of sampling a parameter value $\theta$ in Algorithm 2 is proportional to
\begin{align*}
&\sum_{\ssb}\sum_{\phibn}\pi(\theta)\mathbf{I}_{\{\theta,\tilde{\theta}\}}\pi(\phibn)q(\db|\phibn,\tilde{\db})
\mathbf{I}_{\{||\ssb-\ssbobs||<\epsilon\}}\\
& = \sum_{\ssb}\sum_{\phibn}\pi(\theta,\phibn)\mathbf{I}_{\{\theta,\tilde{\theta}\}}q(\db|\phibn,\tilde{\db})\mathbf{I}_{\{||\ssb-\ssbobs||<\epsilon\}}\\
&\propto \sum_{\ssb}\sum_{\phibn} \pi(\theta,\phibn|\Qb)\mathbf{I}_{\{||\ssb-\ssbobs||<\epsilon)\}}\\
& \propto \pi_\epsilon(\theta|\dbobs,\Qb),
\end{align*}
where the third line follows from the fact that the expression on the second line is the product of the likelihood under the model $\Qb$ and the prior, and therefore it is proportional to the posterior distribution of parameters based on the model $\Qb.$
%
%----------------------------------------------------------------------------------
%
%
%------------------------------------------------------------------
%-------------------------------------------------------------------
%The acceptance region $||\ssbobs-\ssb^*_i||<\epsilon$ used in step 9 of Algorithm 2 corresponds to a uniform kernel on $(\ssb_o-\epsilon,\ssb_o+\epsilon)$ as in the standard ABC approach.
%%
%A non-uniform kernel in the acceptance step generalizes the AABC approach to importance sampling.
%%
%Algorithm 2 might have a low acceptance rate when the prior distribution $\pi(\theta)$ is noninformative and the posterior distribution $\pi_{\epsilon}(\theta|\dbobs,\Qb)$ is peaked with most of its probability concentrated in a small region of the parameter space.
%%
%In that case, an MCMC algorithm might be more efficient than a rejection algorithm to sample the posterior distribution, since the regions with high probability mass can be explored efficiently by an MCMC algorithm \citep{Marjorametal2003}.   
%%
%An AABC by MCMC algorithm (Algorithm 3) and its validation are given in Appendix 3.
%
%\par
%----------------------------------------------------------------------
\section{Applications}
%----------------------------------------------------------------------
In this section, we investigate the inferential performance of AABC approach with two examples.
The following simulation setup is used in both examples. 
%--------------------------------------------------------
\subsection{Simulation study design}\label{sec:simulation}  
We simulated a reference set with $M=10^5$ realizations $\{(\db_1,\theta_1),(\db_2,\theta_2),...,(\db_{10^5},\theta_{10^5})\},$ by first generating $\theta_i\sim \pi(\theta)$ and then simulating a data set $\db_i\sim \Pbsub_{\theta_i}.$ 
We then sampled 1000 pairs $(\db_i,\theta_i)$ from the reference set, uniformly at random without replacement. 
Thus, we selected 1000 ``true'' parameter values $\theta_i,$ along with corresponding test data sets $\db_i$ generated under each value $\theta_i$ from the model $\Pbsub_{\theta_i}$. 
Further, we built the sets $\Zb,$ with $m=10^2, 5\times 10^2, 10^3, 5\times 10^3, 10^4,5\times 10^4, 10^5$ by sampling the reference set uniformly at random without replacement for $m<10^5,$ and taking all the realizations in the reference set for $m=M=10^5.$
The sample size $n$ of the data is described in each relevant example.
\par
%--------------------------------------------------------------------------------------------
On each test data set, we performed AABC by rejection sampling (Algorithm 2) using each set $\Zb.$
In example 1, where our goal is to compare the performance of the AABC and ABC approaches, we performed ABC analyses by rejection sampling (Algorithm 1) using the same sets $\Zb.$
For all analyses, we obtained a sample from the joint posterior distribution of the parameter vector $\theta$ by accepting the parameter vector values that generated data whose summary statistics were in the top $1$ percentile with respect to the statistics calculated from the test data set, in the sense of equation \ref{eq:euclideandistance}. 
Compared to the approach of fixing the $\epsilon$ cutoff, accepting parameter vectors that generate data whose summary statistics are in a top percentile has the advantage that a desired number of samples from the posterior is always obtained given a total fixed number of proposed parameter values.
This approach is often preferred by ABC practitioners and is convenient in our case for comparing ABC and AABC. 
\par
%---------------------------------------------------
We assessed the accuracy of the posterior samples for each component of the parameter vector $\theta$ separately, using the root sum of squared error for standardized parameter values accepted in the posterior sample.
For a generic scalar parameter $\alpha,$ the root sum of squared errors is given by $\textrm{RSSE}=(1/r)\sqrt{\sum_{j=1}^{r}(\alpha_j-\alpha_T)^2/\textrm{Var}(\mathbf{\alpha})},$
where $\mathbf{\alpha}=(\alpha_1,\alpha_2,...,\alpha_r)$ are $r$ accepted values in the posterior sample, $\alpha_T$ is the true parameter value, and $\textrm{Var}(\mathbf{\alpha})$ is the variance of the set of $r$ values.
We report the mean RSSE over 1000 test data sets as $\textrm{RMSE}=(1/1000)\sum_{i=1}^{1000}\textrm{RSSE}_i$ (see \citet{NunesBalding2010}).
\par
%---------------------------------------------------------------------
\subsection{Example 1: The strength of balancing selection in a multi-locus $K${\em-}allele model}\label{sec:example1}
%--------------------------------------------------------------------
In this section, we consider inference from the stationary distribution of allele frequencies in the diffusion approximation
to a Wright-Fisher model with symmetric balancing selection and mutation \citep{Wright1949}.
If we let $a_i>0,$ with $i=1,2,...,K,$ and $\sum_{i=1}^{K}a_i=1,$ and denote the frequency of allelic type $i$ in the population at a genetic locus, the joint probability density function of allele frequencies $x=(a_1,a_2,...,a_K)$ is $f(x|\sigma, \mu)= c(\sigma,\mu)^{-1}\exp(-\sigma\sum_{i=1}^{K}a_i^2)\prod_{i=1}^{K}a_i^{\mu/K-1}.$
Parameters $\sigma$ and $\mu$ determine the population-scaled strength of balancing selection and the mutation rate, respectively. 
A data set of observed allele frequencies is a random sample of $n$ draws from the population frequencies $f(x|\sigma,\mu).$
\par
%-------------------------------------------------
ABC methods are well-suited for inference from this model for three reasons.
First, the statistics $\sum_{j=1}^{K}a_j^2$ and $-\sum_{j=1}^{K}\log a_j$ are jointly sufficient for parameters $\sigma$ and $\mu,$ and no information loss occurs in dimension reduction to the summary statistics. %\citep{BuzbasJoyce2009}.
Second, the parameter-dependent normalizing constant $c(\sigma,\mu)$ is hard to calculate, and performing likelihood-based inference on $\sigma$ and $\mu$ is therefore difficult.
Third, a method specifically designed to simulate data sets from $f(x|\sigma,\mu)$ is readily available \citep{Joyceetal2012}, and performing ABC is therefore straightforward.
For simplicity, we assume 100 loci with the same true parameter values, each with $K=4,$ and that the allele frequencies at each locus are independent of the allele frequencies at other loci.
Thus, the joint probability density function of allele frequencies for 100 loci is equal to the product of probability density functions across loci. 
We choose uniform prior distributions, on $(0.1,10)$ for the mutation rate $(\mu),$ and on $(0,50)$ for the selection parameter $(\sigma)$.
\par
\begin{figure}
\begin{center}
\includegraphics[height=7.6cm]{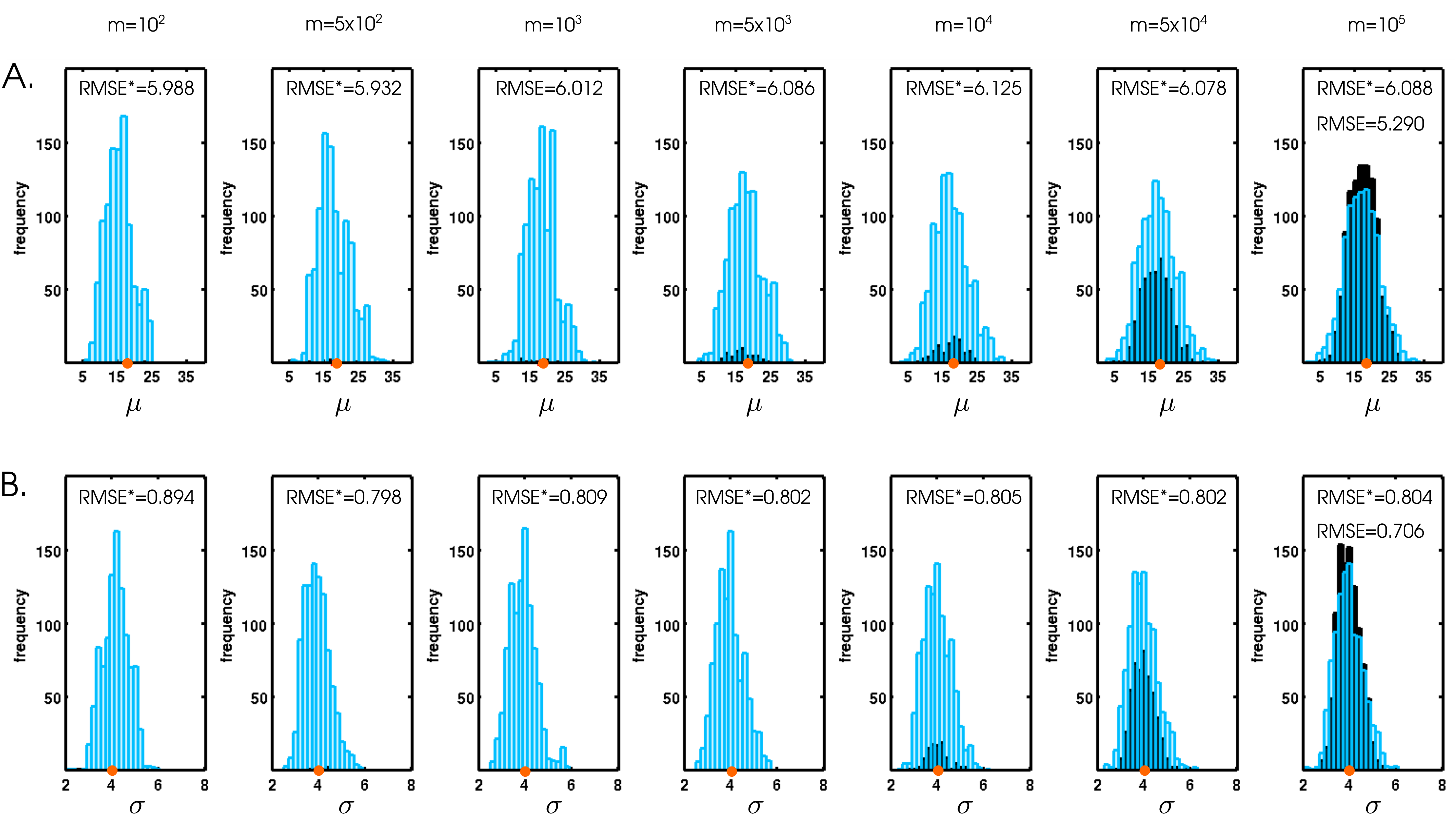}
\caption[]{Inference on the strength of balancing selection. The figure shows the marginal posterior distributions of parameters $\mu$ (A), and $\sigma$ (B) of example 1 obtained with ABC by rejection (black) and with AABC by rejection (blue).  The number $m$ of data sets simulated from the mechanistic model for each analysis performed by AABC and ABC appears at the top of each column. The red dot on the x-axis is the true value of the parameter, equal in all plots. RMSE$^*$ values in each plot are from AABC analyses, averaged over 1000 test data sets. RMSE values in the last column are from corresponding ABC analyses.}
\label{fig:4}
\end{center}
\end{figure}

%----------------------------------------------------------------------------
{\em Results.}
Posterior samples model parameters $(\sigma,\mu)$ obtained by ABC and AABC using a typical data set are given in Figure \ref{fig:4}. 
In analyses with $m=10^2,5\times 10^2, 10^3$ or $5\times 10^3$ simulated data sets, few samples are accepted with ABC, and thus, little mass is observed in ABC histograms (black). 
For small $m,$ ABC does not produce an adequate sample size from the posterior distribution of parameters.
%(on average $1,10,$ and $50$ respectively, based on $1\%$ acceptance rate  for columns 1,2,3).
%
AABC, however, produces a posterior sample of size $10^3$ for any $m,$ because $10^5$ data sets are simulated from the non-mechanistic model (Algorithm 2, steps 5, 6, 7) and the top 1 percentile are accepted as belonging to the approximate posterior distribution.
The histograms obtained by AABC recover the true value reasonably well (Figure \ref{fig:4}).
The RMSE values in AABC procedures are approximately constant with increasing $m.$ 
For $m=10^2,5\times 10^2, 10^3, 5\times 10^3, 10^4, 5 \times 10^4,$ and $10^5$ simulated data sets, the RMSE values for parameter $\mu$ are 5.988, 5.932, 6.012, 6.086, 6.125, 6.078, and 6.088 respectively, close to the RMSE of 5.290 obtained by a standard ABC approach using $M=10^5$ simulated data sets from the mechanistic model.
The RMSE values in the last column of Figure \ref{fig:4} show that an AABC approach produces posterior samples that have on average greater variance than posterior samples obtained from ABC with the same large number of realizations.
Here, greater variance in posterior samples obtained by AABC is a result of simulating data sets in AABC by resampling the observed data values that are found only in the $m$ realizations in $\Zb.$ 
Consider two parameter values $\theta^*_1$ and $\theta^*_2$ for which data sets $\db^*_1$ and $\db^*_2$ are simulated in the AABC approach by steps 5, 6, 7 of Algorithm 2 such that the parameter value $\tilde{\theta}\in\Zb$ closest to both $\theta^*_1$ and $\theta^*_2$ is the same value.
The data sets $\db^*_1$ and $\db^*_2$ can include only the data values observed in $\tilde{\db}$ of the pair $(\tilde{\db},\tilde{\theta})\in\Zb.$ 
On average, $\db^*_1$ and $\db^*_2$ share more observations in common than two data sets simulated from the respective mechanistic models $\Pbsub_{\theta^*_1}$ and $\Pbsub_{\theta^*_2}.$
Therefore, each data set simulated in the AABC approach using $\Qb$ is expected to be less able to distinguish between different parameter values than the independent data sets simulated in the ABC approach using $\Pb.$ 
This situation results in relatively flat likelihoods and hence posterior samples with larger variance.
\par
%-----------------------------------------------------------------------
%-------------------------------------------------------------------
%--------------------------------------------------------
\subsection{Example 2: Admixture rates in hybrid populations}
Models in which hybrid populations are founded by, and receive genetic contributions from, multiple source populations are of interest in describing the demographic history of admixture.  
Stochastic models including admixture often result in likelihoods that are difficult to calculate, and statistical methods capable of performing inference on admixture rates have received much attention for their implications on topics ranging from human evolution to conservation ecology \citep{Falushetal2003, Tangetal2005, BuerkleLexer2008}.
Here, we consider inference on admixture rates from a mechanistic model of \citet{VerduRosenberg2011}.
We use reported estimates of individual admixture as data.
\par
\begin{figure}
\begin{center}
\includegraphics[height=5.5cm]{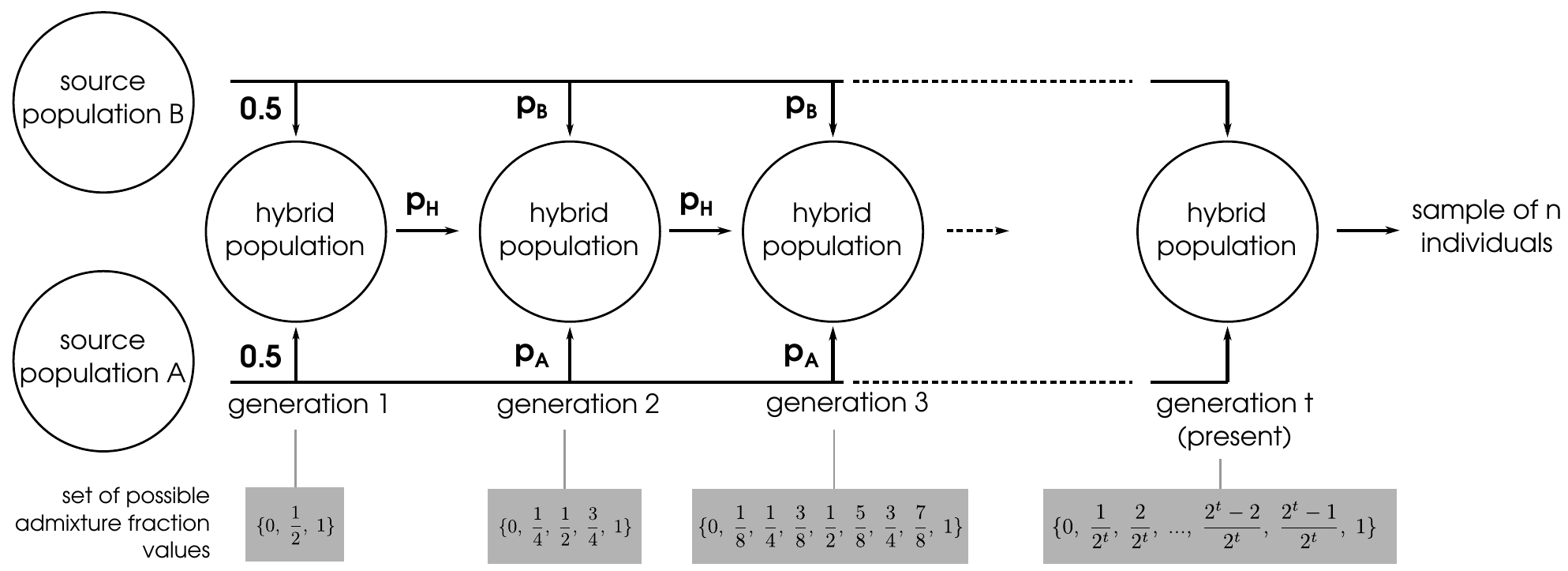}
\caption[]{The admixture model of example 2.}
\label{fig:5}
\end{center}
\end{figure}

%-------------------------------------------------------------------
We consider a model of admixture for a diploid hybrid population of constant size $N,$ founded at some known $t$ generations in the past with contributions from source populations A and B.
We follow the distribution of admixture fractions of individuals in the hybrid population at a given genetic locus.
Each generation, the admixture fraction for each individual in the hybrid population is obtained as the mean of the admixture fractions of its parents.
The parents are chosen independently of each other, from source population A, source population B, or the hybrid population of the previous generation with probabilities $p_A,p_B,$ and $p_H,$ respectively ($p_A+p_B+p_H=1$).
In the special case of the founding generation, $p_H=0,$ and we assume $p_A=p_B=0.5.$ 
Individuals from source populations A and B are assigned admixture fractions of $1$ and $0$ respectively.  
For example, if both parents of an individual in the hybrid population of the founding generation are from source population A, that individual has admixture fraction 
$(1+1)/2=1.$ 
If both parents are from population 2, the admixture fraction is $(0+0)/2=0,$ and if one parent is from population 1 and the other is from population B, then the admixture fraction is $(1+0)/2=0.5.$  
%
%Each generation after founding, $2N$ parents are randomly sampled independently of each other and paired, where the probability of a parent belonging to source population 1, 
%source population 2 and the hybrid population are $s_1,s_2$, and $h$ respectively, $(s_1+s_2+h=1).$ 
%
The distribution of the admixture fraction in the hybrid population is propagated in this manner for $t$ generations until the present, in which a sample of $n$ individuals is obtained from the resulting distribution (Figure \ref{fig:5}).
Our goal is to estimate the admixture rates $(p_A,p_B,p_H),$ given the individual admixture fractions estimated from observed genetic data. 
%
%We use the software STRUCTURE \citep{Pritchardetal2000} to estimate the admixture fraction value of each individual from the genetic data and assume that these estimated values are parametric values of admixture fractions.
%
%A variety of statistical methods are available to estimate individual admixture fractions (see \citet{Tangetal2006}).  
%
%%----------------------------------------------
%%------------------------------------------------------
\par
%------------------------------------------------------------------------------
We apply the AABC approach using individual admixture fractions from $n=604$ individuals from Central African Pygmy populations reported by \citet{Verduetal2009}, with an assumed constant population size of $N=10^4.$
This assumption differs slightly from  the original model in \citet{VerduRosenberg2011} in that a finite population size is assumed, so that only $10^4$ admixture fraction values are allowed in the population at any given generation.  
We assume that an admixture event with contributions from two ancestral source populations started at the mean estimate of $t=771$ generations ago \citep{Verduetal2009} with a generation time of 25 years, and that it continued until the present.
Source population A refers to an ancestral Pygmy population, and source population B refers to an ancestral non-Pygmy population.
The feature of this model relevant to our method is the computational intractability of simulating data sets. 
For each set of parameter values $(p_A,p_B,p_H)$ simulated from the priors, the distribution of admixture fractions is discrete on a support of a number of admixture fraction values that doubles each generation, and this distribution evolves for 771 generations.
A random sample of admixture fraction values comparable to the values calculated from the observed data set is obtained from the distribution of the present generation.
Simulating a large number of data sets under this model with such a large number of generations is computationally infeasible, and standard ABC is impractical.
We thus perform AABC by rejection (Algorithm 2) using $m=10^4$ realizations from this model. 
We assume a Dirichlet prior with hyperparameters $(1,1,1)$ on parameters $(p_A,p_B,p_H).$
\par

%------------------------------------------------------------------------------
We also assessed the contribution of the approximations on the parameter and model spaces in the AABC approach to the RMSE separately, with a simulation study using a small number of generations ($t=30$), where simulating data sets from the mechanistic model is feasible.
First, we performed AABC with rejection as in Algorithm 2 with 1000 ``true'' data sets using $m=10^2, 5\times10^2,10^3, 5\times10^3,10^4, 5\times10^4,$ and $10^5$ realizations from the model, and we calculated the RMSE for $p_A,p_B,$ and $p_H$ over 1000 ``true'' data sets as described in Section \ref{sec:simulation}.
This AABC analysis includes error due to approximations on the parameter space and on the model space.
Second, we performed an AABC analysis with the same set of $m$ realizations, by including the error only due to the approximation on the parameter space.
We achieved this by running Algorithm 2 up through step 5, and then simulating data sets from the mechanistic model by substituting steps 6 and 7 of Algorithm 2 with step 2 of Algorithm 1, the standard ABC approach by rejection.
By this substitution, all data sets are simulated from the mechanistic model, but each data set is obtained using a parameter vector $(\tilde{p}_A,\tilde{p}_B,\tilde{p}_H)$ found in step 5 of Algorithm 2.
In this procedure, the error due to the approximation on the model space is eliminated, because data sets are simulated from the correct mechanistic model and not by resampling from the available realizations in $\Zb$.
However, this procedure includes error due to the approximation on the parameter space, because each data set is simulated not under the correct proposed parameter value, but under the parameter value $(\tilde{p}_A,\tilde{p}_B,\tilde{p}_H),$ the closest value to the correct proposed value that can be found in $\Zb.$
We compared the RMSE of the AABC procedure involving the approximation on both the parameter and model spaces and the RMSE of the AABC procedure involving only the approximation on the parameter space to the RMSE obtained from a standard ABC approach.
For these two AABC procedures, we also compared the percent excess in RMSE, defined as the ratio of the absolute difference in RMSE of the AABC and standard ABC approaches to the RMSE of the standard ABC approach, expressed as a percent.  
%-----------------------------------
\par
\begin{figure}
\begin{center}
\includegraphics[height=11cm]{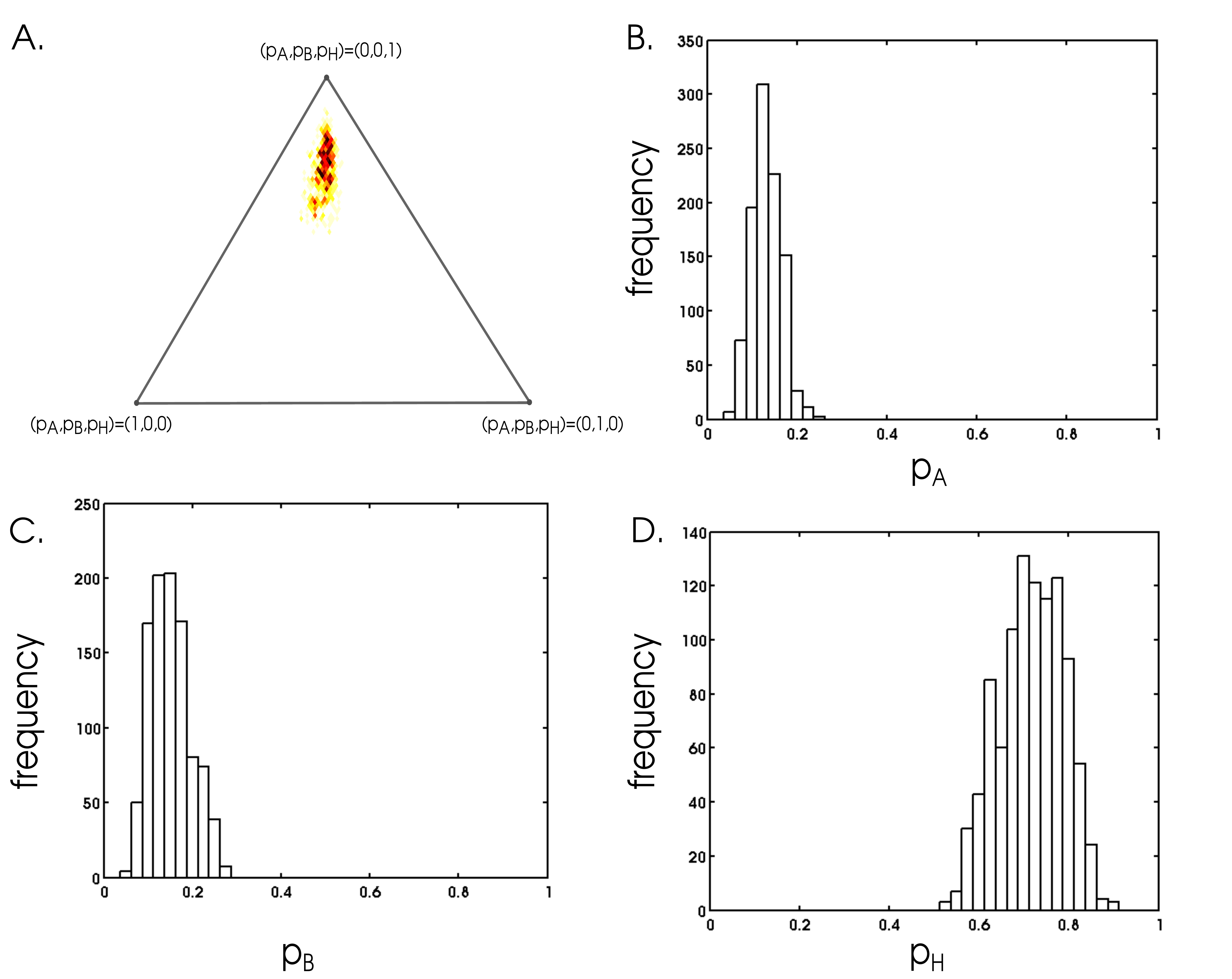}
\caption[]{AABC analysis on the Pygmy data of example 2 with \(m=10^4\) realizations under the mechanistic model. (A) The joint distribution on the unit simplex, with probability mass increasing from white to dark red. (B,C,D) Marginal distributions of \(p_A,p_B,\) and \(p_H.\)}
\label{fig:6}
\end{center}
\end{figure}

%------------------------------------------------------------------------------
{\em Results.}
The individual admixture fractions calculated from the Pygmy data carry substantial information about the admixture parameters $p_A,p_B,$ and $p_H,$ since the joint posterior distribution is concentrated in a relatively small region of the 3-dimensional unit simplex on which $(p_A,p_B,p_H)$ sits (Figure \ref{fig:6}A). The marginal posterior distributions (Figure \ref{fig:6}B, \ref{fig:6}C, and \ref{fig:6}D) have means $p_A=0.151,\; p_B=0.132,$ and $p_H=0.717.$ 
These values are interpreted as contribution of genetic material of 15.1\% from the ancestral Pygmy population (source population A), 13.2\% from the ancestral Non-Pygmy population (source population B), and 71.7\% from the hybrid population to itself at each generation, over $771$ generations of constant admixture. 
%%----------------------------------------------
%%------------------------------------------------------
\par
\begin{figure}
\begin{center}
\includegraphics[height=13cm]{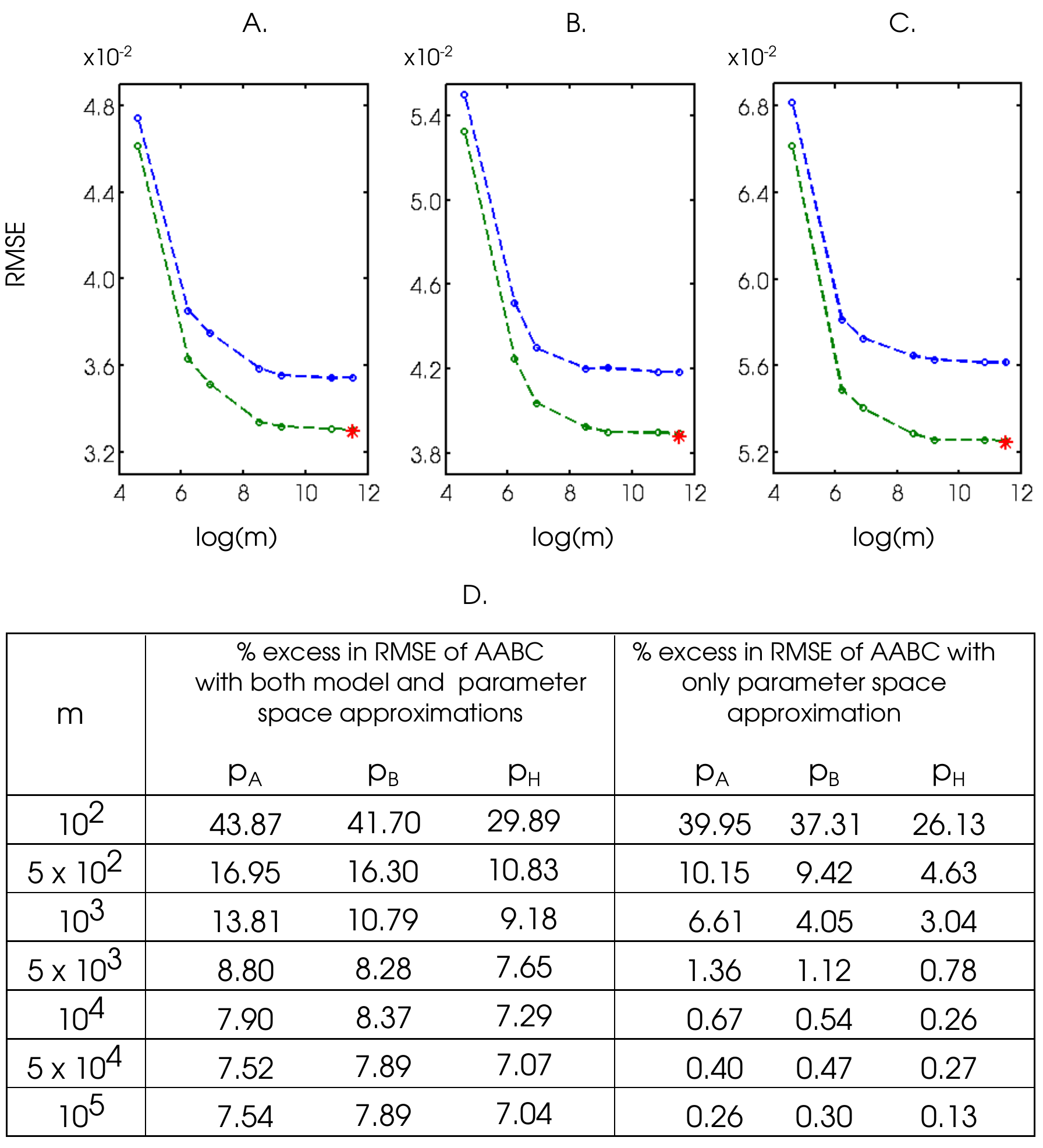}
\caption[]{RMSE in the admixture model. The decrease in RMSE is shown for parameters \(p_A\) (A), \(p_B\) (B), and \(p_H\) (C) with increasing \(m,\) the number of simulated samples from the mechanistic model, for AABC analysis performed with an approximation only on the parameter space (green), and with an approximation on both the parameter space and the model space (blue). The red star in each plot is the RMSE obtained by a standard ABC analysis performed with \(M=10^5\) simulated values. (D) The percent excess in RMSE of the two AABC approaches relative to a standard ABC approach for parameters \(p_A,p_B,\) and \(p_H\).} 
\label{fig:7}
\end{center}
\end{figure}

%------------------------------------------------------------------------------
%
For the simulation study with $t=30$ generations and 1000 ``true data'' sets, the RMSE values from AABC analyses decrease with increasing $m$ (Figure \ref{fig:7}A, \ref{fig:7}B,
\ref{fig:7}C). 
Further, as $m$ increases, the error due to the approximation on the parameter space decreases (Figure \ref{fig:7}D last column), due to the fact that for large $m,$ the difference decreases between the closest parameter value chosen at step 5 of Algorithm 2 and the correct parameter value under which we want to simulate a data set.
In fact, the RMSE from the AABC analysis with $m=10^5$ realizations and approximation only on the parameter space and the RMSE from the standard ABC approach are virtually indistinguishable (Figure \ref{fig:7}A, \ref{fig:7}B, \ref{fig:7}C, red star).  
For $m=10^3,$ the AABC analysis with approximations on the parameter and model spaces has a percent excess RMSE of 13.81\%, whereas AABC analysis including only the approximation on the parameter space has excess RMSE of 6.61\%. 
That is, at $m=10^3,$ approximately half of the excess RMSE in the AABC approach with respect to the standard ABC analysis comes from the error due to the approximation on the parameter space and half arises due to the approximation on the model space.  
%  
%%----------------------------------------------
%%------------------------------------------------------
\section{Discussion}
%-------------------------------------------------------------------------------------- 
Performing likelihood-based inference from statistical models incorporating a multitude of stochastic processes is often challenging due to computationally intractable likelihoods.
In principle, when stochastic processes are complex but a family of parametric statistical models is well-defined, data can be simulated from the model to assess the parameter likelihoods. 
In the last decade, ABC methods have become a standard tool to perform approximate Bayesian inference in subject areas such as ecology and evolution, by exploiting
the idea of simulating many data sets from a model, when such simulations are computationally feasible.
To deliver an adequate sample from the posterior distribution of the parameters, however, ABC requires a large number of simulated data sets, and it might not perform well when only a limited number of data sets can be simulated. 
\par
%--------------------------------------------------------------------------------
In this article, we introduced an approach that extends simulation-based Bayesian inference methods to model spaces in which only a limited number of data sets can be simulated from the model, at the expense of requiring approximations on the parameter and the model spaces.
Our AABC approaches rely on two statistical approximations. 
In our approximation on the parameter space, for each parameter simulated from the prior distribution, we take the closest parameter value available in the set of realizations 
$\Zb$ obtained from the mechanistic model.
This approach has a uniform kernel smoothing interpretation
in the sense that each parameter value in the set $\Zb$ dissects the support of the prior distribution into non-overlapping components such that each interval is mapped to the same parameter value in $\Zb.$ 
Each component then represents the support of a uniform kernel.      
Kernel approximations have an operational role in implementing ABC methods, and a natural future direction for AABC is to improve the accuracy of posterior samples using smooth weighting kernels for the approximation on the parameter space.
\par
%------------------------------------------------------
The approximation on the model space is achieved by assigning Dirichlet probabilities to data points of realizations obtained from the mechanistic model.
This is a variation on the resampling method originally introduced in Rubin's Bayesian bootstrap \citep{Rubin1981}, and therefore, it is an application of Bayesian nonparametric methods.
From this perspective, AABC methods connect standard model-based Bayesian inference on model-specific parameters and Bayesian nonparametric methods within the ABC framework.
\par
%-----------------------------------------------------------------
Our approach of using a non-mechanistic model and Bayesian resampling methods to help perform inference on model-specific parameters of a mechanistic model is a fundamental difference between AABC and existing ABC methods.
ABC performs inference on model-specific parameters of a mechanistic model using a likelihood based purely on the mechanistic model.
AABC instead performs inference on the same model-specific parameters of the mechanistic model as ABC, using a likelihood based on a non-mechanistic model that incorporates a limited number of data sets simulated from the mechanistic model.
Consequently, the model likelihoods used in ABC and AABC are not exactly the same, and the posterior distributions targeted by the two classes of methods are not exaxctly equivalent for finite sample sizes.
The advantage of AABC methods in contrast to pure non-mechanistic modeling approaches (e.g., nonparametric methods) is that AABC can perform inference on the quantities of interest---the model-specific parameters of the mechanistic model.
%----------------------------------------------------------------------
\par
%----------------------------------------------------------
Unlike other ABC methods, the AABC approach delivers a posterior sample of desired size from the joint distribution of parameters for any $m>1.$
This is both a strength and a limitation of AABC.
The strength is that in practice, a researcher can fix $m$ and thus the computation time {\em a priori}, to simulate data from the mechanistic model to obtain a reasonable inference by AABC; other ABC methods may fail to produce an adequate posterior sample in equivalent computation time. 
In our example, for moderate values of $m$ (e.g., $10^3$ to $10^4$) for which standard ABC approaches were unsatisfactory, AABC adequately sampled an approximate posterior distribution.
The limitation is that when $m$ is too small, the posterior sample obtained by AABC can be a distorted representation of the true posterior distribution.
Although in the limit, AABC and ABC are expected to produce similar results, the posterior distribution sampled by an AABC approach is not the correct posterior distribution, because many parameter values simulated from the prior are tested for acceptance based on repeated use of the data values in $m$ realizations, instead of based on data sets simulated independently of each other.  
A future direction is to investigate the relationship between $m$ and the dimensionality of the parameter space to optimize $m$ in producing a given level of accuracy for approximating the true posterior distributions.  
\par
%-----------------------------------------------------
\section*{Acknowledgments}
The authors thank Paul Verdu for helpful discussions on the genetics of Central African Pygmy populations. Support for this research is partially provided by NIH grant R01 GM 081441, NSF grant DBI-1146722, and the Burroughs Wellcome Fund.
%------------------------------------------------------------------
%--------------------------------------------------------------------------

%-----------------------------------------------------------------
%\appendix
\section*{Appendix 1}
We let $k\leq n$ be the number of distinct values $\tilde{x}_1,\tilde{x}_2,...,\tilde{x}_k$ in the data set $\tilde{\db},$ and denote the number of observed $\tilde{x}_i$ by $\tilde{n}_i,$ where $n=\sum_{i=1}^{k}\tilde{n}_i.$   
Then the prior distribution for the probabilities of an AABC replicate data set based on the ABC simulated data set $\tilde{\db}$ is the Dirichlet distribution $\pi(\phibn)=[\Gamma(\sum_{i=1}^k \tilde{n}_i)/\prod_{i=1}^k\Gamma(\tilde{n}_i)] \prod_{i=1}^{k}\phi^{\tilde{n}_i-1}$ with parameters $\tilde{n}_1,\tilde{n}_2,...,\tilde{n}_k.$ 
The special case of the prior proportional to $1$ described in the text is obtained with $k=n,$ when all observations in $\tilde{\db}$ are distinct $(\tilde{n}_1,=\tilde{n}_2=\;\cdots\;=\tilde{n}_n=1)$. 
Our goal is to show that $\lim_{m\rightarrow \infty}\lim_{n\rightarrow \infty}\pi_\epsilon(\theta|\dbobs,\Qb)=\lim_{n\rightarrow\infty}\pi_{\epsilon}(\theta|\dbobs,\Pb).$ 
\par
Recalling equation \ref{eq:4},
\begin{equation}\label{eq:app1}
 \lim_{m\rightarrow \infty}\lim_{n\rightarrow \infty}\pi_\epsilon(\theta|\dbobs,\Qb)= \displaystyle{\lim_{m\rightarrow\infty}\lim_{n\rightarrow \infty}}\frac{1}{C_{\Qb}}\int_{\mathcal{X}}\mathbf{I}_{\{||\ssb-\ssbobs||<\epsilon\}}\left[\int_{\Phi}q(\db|\phibn,\tilde{\db})\pi(\phibn)\;d\phibn\;\mathbf{I}_{\{\theta,\tilde{\theta}\}}\right]\pi(\theta)\; d\db.
\end{equation}
The integral in the brackets is the expectation of $q(\db|\phibn,\tilde{\db}),$ with respect to the prior $\pi(\phibn).$ We let $C={n \choose n_1 \;n_2\; \cdots\; n_k},$ and using the definition of $q(\db|\phibn,\tilde{\db})=C\;\prod_{j=1}^{n}\prod_{i=1}^{n}\phi_i^{\mathbf{I}_{\{x_j=\tilde{x}_i\}}}$ in section \ref{subsec:nonparametric}, and $\pi(\phibn)=[\Gamma(\sum_{i=1}^k \tilde{n}_i)/\prod_{i=1}^k\Gamma(\tilde{n}_i)] \prod_{i=1}^{k}\phi^{\tilde{n}_i-1}$ we get
\begin{equation*}
\int_{\Phi}q(\db|\phibn,\tilde{\db})\pi(\phibn)\;d\phibn=C\;\frac{\Gamma(\sum_{i=1}^{k}\tilde{n}_i)}{\prod_{i=1}^{k}\Gamma(\tilde{n}_i)} \;\prod_{j=1}^{n}\int_{\Phi}\left(\prod_{i=1}^{n}\phi_i^{\mathbf{I}_{\{x_j=\tilde{x}_i\}}}\right)\left(\prod_{i=1}^{k}\phi_i^{\tilde{n}_i-1}\right)\;d\phibn.
\end{equation*}
Here, we have exchanged the order of the product over $j$ with the integral since the expectation of the product of $n$ IID observations in sample $\db$ is equal to the the product of the expectations of observations $x_j.$ 
We label the realized value of the $j$th data point $x_j$ by $(j)$ such that $\prod_{i=1}^{n}\phi_i^{\mathbf{I}_{\{x_j=\tilde{x}_i\}}}=\phi_{(j)},$ and write
\begin{equation}\label{eq:app2}
\int_{\Phi}q(\db|\phibn,\tilde{\db})\pi(\phibn)\;d\phibn=C\;\frac{\Gamma(\sum_{i=1}^{k}\tilde{n}_i)}{\prod_{i=1}^{k}\Gamma(\tilde{n}_i)}\;\prod_{j=1}^{n}\int_{\Phi} \left(\prod_{\substack{i=1\\i\neq(j)}}^{k}\phi_i^{\tilde{n}_i-1}\right)\phi_{(j)}^{\tilde{n}_{(j)}}\;d\phibn.
\end{equation}
Using 
$\int_{\Phi}\frac{\Gamma[(\sum_{i=1,i\neq (j)}^{k}\tilde{n}_i)+\tilde{n}_{(j)}+1]}{[\prod_{i=1,i\neq (j)}^{k}\Gamma(\tilde{n}_i)]\Gamma(\tilde{n}_{(j)}+1)} \; \left(\prod_{i=1,i\neq (j)}^{k}\phi_i^{\tilde{n}_i-1}\right)\phi_{(j)}^{\tilde{n}_{(j)}}\;d\phibn=1$
(p. 487, \citet{Kotzetal2000}), we substitute the integral in equation (\ref{eq:app2}) with the ratio of the gamma functions to get
\begin{align*}
\int_{\Phi}q(\db|\phibn,\tilde{\db})\pi(\phibn)\;d\phibn&=C\;\frac{\Gamma(\sum_{i=1}^{k}\tilde{n}_i)}{\prod_{i=1}^{k}\Gamma(\tilde{n}_i)}\prod_{j=1}^{n}
\frac{\left[\prod_{i=1,i\neq (j)}^{k}\Gamma(\tilde{n}_i)\right]\Gamma(\tilde{n}_{(j)}+1)}{\Gamma[(\sum_{i=1,i\neq (j)}^{k}\tilde{n}_i)+\tilde{n}_{(j)}+1]}\\
&=C\;\prod_{j=1}^{n}\frac{\Gamma(n)}{\Gamma(\tilde{n}_{(j)})}\frac{\Gamma(\tilde{n}_{(j)}+1)}{\Gamma(n+1)}=C\;\prod_{j=1}^{n}\left(\frac{\tilde{n}_{(j)}}{n}\right).
\end{align*}
Substituting $C\;\prod_{j=1}^{n}\left(\frac{\tilde{n}_{(j)}}{n}\right)$ for the integral in brackets in equation (\ref{eq:app1}), we have
\begin{align}
\nonumber
\lim_{m\rightarrow \infty}\lim_{n\rightarrow \infty}\pi_\epsilon(\theta|\dbobs,\Qb)&=\displaystyle{\lim_{m\rightarrow\infty}\lim_{n\rightarrow \infty}}\frac{1}{C_{\Qb}}\int_{\mathcal{X}}\mathbf{I}_{\{||\ssb-\ssbobs||<\epsilon\}}\;C\;\prod_{j=1}^{n}\left(\frac{\tilde{n}_{(j)}}{n}\right)\;\mathbf{I}_{\{\theta,\tilde{\theta}\}}\pi(\theta)\; d\db\\
\label{eq:exchangelimit}
&=\frac{\displaystyle{\lim_{m\rightarrow\infty}\lim_{n\rightarrow \infty}}\int_{\mathcal{X}}\mathbf{I}_{\{||\ssb-\ssbobs||<\epsilon\}}\;C\;\prod_{j=1}^{n}\left(\frac{\tilde{n}_{(j)}}{n}\right)\;\mathbf{I}_{\{\theta,\tilde{\theta}\}}\pi(\theta)\; d\db}{\displaystyle{\lim_{m \rightarrow \infty }\lim_{n\rightarrow \infty}}C_{\Qb}}.
\end{align}
\par
We apply the dominated convergence theorem to exchange the limits in $n$ and the integrals in the numerator and denominator of equation (\ref{eq:exchangelimit}).
The assumptions of the theorem are satisfied as follows:
1) The integrand in equation (\ref{eq:exchangelimit}) is bounded: The indicator functions are bounded by 1, the ratios $(\tilde{n}_{(j)}/n),$ where $n_{(j)}\leq n$ are bounded by 1, and the prior $\pi(\theta)$ is bounded by assumption.
2) $\lim_{n\rightarrow \infty}(\tilde{n}_{(j)}/n)$ converges pointwise to the probability of $x_{(j)}$ under $\tilde{\theta}$ and the model $\Pbsub_{\tilde{\theta}},$ given by 
$p(x_{(j)}|\tilde{\theta}),$ by the frequency interpretation of probability.
Exchanging the limits in $n$ and the integrals, and using $\lim_{n\rightarrow \infty}(\tilde{n}_{(j)}/n)=p(x_{(j)}|\tilde{\theta}),$
\begin{align}
\nonumber
 \lim_{m\rightarrow \infty}\lim_{n\rightarrow \infty}\pi_\epsilon(\theta|\dbobs,\Qb)&=\frac{\displaystyle{\lim_{m\rightarrow\infty}}\int_{\mathcal{X}}\mathbf{I}_{\{||\ssb-\ssbobs||<\epsilon\}}\prod_{j=1}^{k}\left[p(x_{(j)}|\tilde{\theta})\right]^{n_{(j)}}\;\mathbf{I}_{\{\theta,\tilde{\theta}\}}\pi(\theta)\;d\db}{\displaystyle{\lim_{m \rightarrow \infty}}C_{\Pbsub_{\tilde{\theta}}}}\\
\label{eq:app4}
&=\frac{\displaystyle{\lim_{m\rightarrow\infty}}\int_{\mathcal{X}}\mathbf{I}_{\{||\ssb-\ssbobs||<\epsilon\}}p(\db|\tilde{\theta})\;\mathbf{I}_{\{\theta,\tilde{\theta}\}}\pi(\theta)\;d\db}{\displaystyle{\lim_{m \rightarrow \infty}}C_{\Pbsub_{\tilde{\theta}}}},
\end{align}
where (\ref{eq:app4}) follows by the definition of the joint distribution $p(\db|\tilde{\theta})=\prod_{j=1}^{k}\left[p(x_{(j)}|\tilde{\theta})\right]^{n_{(j)}}.$
\par
We now apply the dominated convergence theorem a second time to exchange the limits in $m$ and the integrals on $\mathcal{X}$. Again, the assumptions of the dominated convergence theorem are satisfied since  the integrand in (\ref{eq:app4}) is a sequence in $m$ of bounded functions, and as $m\rightarrow \infty,$ $\tilde{\theta}\rightarrow \theta,$ and $p(\db|\tilde{\theta})\rightarrow p(\db|\theta).$ 
We get
\begin{equation*}
\lim_{m\rightarrow \infty}\lim_{n\rightarrow \infty}\pi_\epsilon(\theta|\dbobs,\Qb)=\frac{1}{C_{\Pb}}\int_{\mathcal{X}}\mathbf{I}_{\{||\ssb-\ssbobs||<\epsilon\}}p(\db|\theta)\pi(\theta)\;d\db=\displaystyle{\lim_{n\rightarrow \infty}}\pi_{\epsilon}(\theta|\dbobs,\Pb)
\end{equation*}
which shows that AABC posterior converges to the ABC posterior as the sample size $n$ and the simulated number of data sets $m$ increase.
%--------------------------------------------------------------------------------
\end{document}